\documentclass[usenatbib]{mn2e}
\usepackage{graphicx}
\newcounter{paper}
\begin{document}

\title[H$\alpha$ survey of early-type cluster galaxies] 
{An H$\alpha$ survey of cluster galaxies \Roman{paper}: cluster -- field 
comparison for early-type galaxies}

\author[C. Moss and M. Whittle]
{C.~Moss$^1$ and M.~Whittle$^2$ \\ 
$^1$Astrophysics Research Institute, Liverpool John Moores
University, Birkenhead CH41 1LD \\
$^2$Department of Astronomy, University of Virginia, Charlottesville,
VA 22903, USA}

\maketitle

\begin{abstract}
We have extended our H$\alpha$ objective prism survey of 8
low-redshift clusters (viz. Abell 262, 347, 400, 426, 569, 779, 1367
and 1656) to include a complete sample of early-type galaxies within
1.5 Abell radii of the cluster centres.  Of the 379 galaxies surveyed,
3\% of E, E--S0 galaxies, 6\% of S0s and 9\% of S0/a galaxies were
detected in emission.  From a comparison of cluster and supercluster
field galaxies, we conclude that the frequency of emission-line
galaxies (ELGs; $W_{\lambda} \ge 20$\AA) is similar for field and
cluster early-type galaxies.  A similar result has previously been
obtained for galaxies of types Sa and later.  Together, these results
confirm the inference of Biviano et al. that the relative frequency of
ELGs in clusters and the field can be entirely accounted for by the
different mix of morphological types between the differing
environments, and that, for galaxies of a given morphological type,
the fraction of ELGs is independent of environment.  Detected emission
is classified as `compact' or `diffuse', identified as circumnuclear
starburst or AGN emission and disk emission respectively.  By
comparing spectroscopic data for cluster early-type ELGs with data for
field galaxies from the Palomar Spectroscopic Survey of nearby
galactic nuclei, we demonstrate there is modest evidence for an
enhancement of compact HII emission relative to AGN emission in the
early-type cluster ELGs as compared to the field.  For the cluster
early-type galaxies, compact HII emission correlates strongly with a
disturbed morphology.  This suggests that, as for later type cluster
galaxies, this enhanced compact HII emission can readily be explained as an
enhancement of circumnuclear starburst emission due to gravitational
tidal interactions, most likely caused by sub-cluster merging and
other on-going processes of cluster virialisation.
\end{abstract}

\begin{keywords}
surveys -- galaxies: clusters:general -- galaxies: elliptical and lenticular,cD
 -- galaxies: evolution -- galaxies: interactions -- galaxies: starburst
\end{keywords}

\section{Introduction}
\label{intro}

Clusters of galaxies are sites of strong morphological evolution of
disk galaxies. While the fraction of ellipticals in clusters remains
relatively unchanged with redshift, the fraction of spirals is 2--3
times larger in intermediate redshift clusters as compared to the
present, with a corresponding decrease in the S0 population (Dressler
1980; Dressler et al. 1997). This dramatic change in the disk galaxy
population since $z \sim 0.5$ suggests that processes associated with
cluster virialisation cause the transformation of spirals to S0s.

Although a variety of mechanisms have been proposed to explain the
morphological change of cluster disk galaxies, there is growing
evidence to suggest that gravitational tidal effects are the
predominant mechanism for these transformations. Intermediate redshift
clusters have an unexpectedly high proportion of galaxies with unusual
morphology, suggestive of merging and tidally interacting systems
(e.g. Lavery \& Henry 1988; Thompson 1988; Lavery, Pierce \& McClure
1992; Dressler et al. 1994; Couch, Ellis \& Sharples 1994; Oemler,
Dressler \& Butcher 1997).  Furthermore, the dynamically interacting
galaxies seem to be responsible for most of the galaxies that show
spectroscopic signs of starbursts (Oemler et al. 1997), as may be
expected from the consequences of tidal interactions (e.g. Barton et
al. 2000). These tidal effects and gravitational interactions are able
to transform spirals to S0s by gas stripping (e.g. Valluri \& Jog 1991);
by the loss of angular momentum of the disk gas, igniting a powerful
central starburst which can assist in the formation of a bulge (e.g. 
Hernquist \& Mihos 1995; Barnes \& Hernquist 1996); by tidal heating
of the disk stabilising it against gravitational instability and 
suppressing subsequent star formation (e.g. Gnedin 2003b); by
truncation of the halo, halting any further infall of cold gas (e.g.
Merritt 1984; Gnedin 2003b); and by minor merger of two spirals to
form an S0 (e.g. Bekki 1998). 

The correlation of star formation rate and local projected density
which holds for galaxies up to several virial radii from the centres
of clusters (e.g. Kodama et al. 2001, 2003; Lewis et al.  2002; Gomez
et al. 2003) suggests that pre-processing due to tidal forces in
galaxy groups is a significant means to change galaxy morphology
(e.g. Zabludoff \& Mulchaey 1998).  However, recent numerical
simulations demonstrate that the tidal forces associated with on-going
cluster formation and sub-cluster merging also play a major role in
the morphological transformation of cluster disk galaxies (e.g. Bekki
1999; Gnedin 1999, 2003a, 2003b).
 
Changes in cluster disk galaxy morphology were dramatic in the past.
However it is important to consider whether such changes are
continuing in clusters at the present epoch.  If morphological
transformations of cluster disk galaxies are occuring at the present,
albeit on a reduced scale, the processes involved may be studied in
much greater detail than is easily possible at higher redshifts.  In
fact, there is growing evidence that activity similar to that seen in
distant clusters is still on-going at the present epoch.  In previous
work (Moss \& Whittle 1993; Moss, Whittle \& Pesce 1998; Moss \&
Whittle 2000), we have shown that the residual spiral population of
low-redshift clusters shares characteristics of the more abundant
spiral population in higher redshift clusters. There is an enhancement
of circumnuclear starburst emission in cluster spirals as compared to
their counterparts in the field. Moreover, a high proportion of
the spirals in low-redshift clusters (e.g. $\sim$ 40\% in Coma) have a
distorted morphology, strongly correlated with circumnuclear starburst
emission, typical of tidally disturbed systems. Similarly Caldwell \&
Rose (1997) in a study of 5 nearby clusters, conclude that 15\% of
early-type galaxies show signs of on-going or recent star formation,
characteristic of starburst or post-starburst galaxies, and that the
frequency of such galaxies is enhanced as compared to the field.
Further evidence that tides and interactions are not solely the 
provenance of intermediate redshift clusters is provided by 
Conselice and Gallagher (1999) who show that distorted, interacting
and fine-scale substructures are common in low-redshift cluster galaxies.   

Our previous work was an H$\alpha$ survey of spirals (types Sa and
later) in 8 low-redshift clusters, which we used to compare the
incidence of star formation in clusters and the field.  In the present
paper we use the same data set to complete the survey for early-type
galaxies, and compare emission for these galaxies between clusters and
the field. Such a comparison is of interest for several reasons.
Firstly, previous such comparisons have generally concluded that
emission is reduced in cluster early-type galaxies as compared to
similar galaxies in the field.  However these earlier results have
been cast in doubt due to the discovery of a previously unrecognised
selection effect (cf. Biviano et al. 1997). Our survey data is
expected to be free of this selection effect, and we are able to
provide an unbiased comparison of emission in cluster and field
early-type galaxies (see discussion in section \ref{scfcomp} below).
Secondly, previous results for low-redshift clusters (cf. Caldwell et
al. 1993; Caldwell
\& Rose 1997), lead to an expectation of an enhancement of starbursts
in the cluster early-type population as compared to the field. Using
the present survey data, we can attempt to confirm this
prediction. And finally, data for detected early-type and late-type
emission-line galaxies (ELGs) may be combined to gain further insight
into processes affecting cluster galaxies particularly from the
kinematical properties of the various ELG subgroups.  This last
investigation is the topic for a subsequent paper (Moss 2004, in
preparation).

Identifications of ELGs from combined H$\alpha$ + [NII] emission for
early-type galaxies from the H$\alpha$ survey data have already been
published for Abell 1367 (cf. Moss, Whittle \& Pesce 1998) In section
\ref{opsurvey} we list the early-type galaxies surveyed, and present
ELG identifications for the remaining 7 clusters. A comparison of
emission in field and cluster early-type galaxies is made in section
\ref{scfcomp}.  In section \ref{emmech} we investigate possible enhancement 
of starburst emission in cluster early-type galaxies.  Conclusions are given 
in section \ref{concl}.

\section{Objective prism survey of early-type cluster galaxies}
\label{opsurvey}

\subsection{Survey sample}
\label{ssample}

An objective prism survey for combined H$\alpha$ + [NII] emission from
cluster galaxies in 8 low-redshift clusters (viz. Abell 262, 347, 400,
426, 569, 779, 1367 and 1656) has been undertaken using the 61/94-cm
Burrell Schmidt telescope on Kitt Peak.  The survey technique and
methods, and the plate material used have been described in detail in
previous papers \citep[][ Papers I--IV
respectively]{mwi,mw93,mwp,mw00}. For convenience, a brief summary of
these details is given here.  For each field, two plates were taken
using hypersensitised IIIaF emulsion, with an emulsion/filter
combination giving a $\sim$ 350\AA\/ bandpass centred on 6655\AA\/
with a peak sensitivity of $\sim$ 6717\AA.  All plates were taken in
conditions of good seeing and good transparency, and an H$\alpha$
detection was accepted only if the galaxy was independently detected
on both plates.  Previous work has shown that the approximate
detection limit of H$\alpha$ + [NII] emission is an equivalent width,
$W_{\lambda}$ $\simeq$ 20\AA\/.

The initial survey list comprised all 727 CGCG galaxies within 
a radial distance from the cluster centre, $r \le 1.5r_{\rm A}$, where
$r_{\rm A}$ is the Abell radius 
\cite{ab58}. In addition, 79 CGCG
galaxies (all in Abell 1367, except one in Abell 400) 
which lie in the region, 
$1.5r_{\rm A} < r \le 2.6r_{\rm A}$, were included in the survey.
Adopted values of the Abell radius for each cluster,
and plate boundaries of the survey plate material are given in Papers
III and IV.  Of the 806 CGCG galaxies in the initial list, 37 are
double systems. The components of these were surveyed separately,
giving a total of 843 galaxies.
 
Galaxy types according to the revised de Vaucouleurs system (de
Vaucouleurs 1959, 1974) were obtained for all galaxies in the
initial survey list. Types were either taken from the UGC
\citep{nil73} or determined by inspection from a variety of Schmidt
plate material.  Details of the type classification procedure are
given in Papers III and IV.  Types for galaxies in Abell 1367 are
given in Paper III, and for galaxies of types Sa and later for the
remaining clusters in Paper IV.

Papers III and IV list types and H$\alpha$ + [NII] emission detection
for 460 galaxies, which are mainly of types Sa and later.  The
present paper completes the survey by listing types and H$\alpha$ +
[NII] emission detection for the remaining 383 galaxies of the survey,
which are predominantly of types S0/a and earlier.

Combining the data from this paper with those for Papers III and IV,
there is a total of 95 galaxies omitted from the survey due to plate
defects (68), or a velocity $\ge$ 12000 km $s^{-1}$ (27).  Thus the
final total of surveyed galaxies is 748.

\subsection{Emission detection}
\label{edetection}

In Table \ref{tsurvey} we give galaxy types for the remaining 383
galaxies of the initial survey list, together with the heliocentric
velocity taken from the NASA Extragalactic Database (NED).  With the
exception of 8 galaxies in Abell 262 typed as spirals and one galaxy
in Abell 347 typed as `peculiar', accidentally overlooked in previous
work, the remaining 374 galaxies are all of types S0/a and earlier, or
are untyped (type class `$\ldots$').  Of the galaxies listed in Table
1, 30 could not be surveyed due to plate defects (these galaxies are
listed in the Notes to the Table), or due to a velocity $\ge$ 12000 km
$s^{-1}$, since in the latter case any H$\alpha$ emission is
redshifted beyond the sensitivity limit of the plate.  Thus there is a
total of 353 surveyed galaxies.

Of these 353 predominantly early-type galaxies, 28 galaxies were
detected in emission.  The emission-line galaxies (ELGs) are listed in
Table \ref{gsurvey}.  For galaxies in Table 2 which are untyped (type
class `...'), we list additional type information from NED, where
available. The visual classification of the detected emission
according to visibility (S -- strong; MS -- medium-strong; M --
medium; MW -- medium-weak; and W -- weak) and concentration (VC --
very concentrated; C -- concentrated; N -- normal; D -- diffuse; and
VD -- very diffuse) is given according to the scheme used in previous
work (cf. Papers I--IV).  Similarly, for the subsequent analysis, we
choose binary ranks for the H$\alpha$ appearance, yielding two
parameters: {\it compact} emission (concentration classes VC, C or N);
and {\it diffuse} emission (concentration classes D or VD).  Notes on
individual objects are appended to the Table.

In Table \ref{etypes} we summarise emission detection frequency with
morphological type.  In order to approximate a volume-limited sample
(cluster galaxies and galaxies proximate to the cluster in the
supercluster field), we restricted the sample to galaxies with
velocities within 3$\sigma$ of the cluster mean.  For each
morphological type class, the Table lists the total sample number
($n_{t}$), the numbers of detected galaxies with compact and diffuse
emission ($n_{e,c}$ and $n_{e,d}$ respectively), and the overall
percentage of galaxies detected in emission ($p_{e}$).  Corresponding
values are also given when NED types, where available, have been
included for galaxies with undetermined types ($n^\prime_t$,
$n^\prime_{e,c}$, $n^\prime_{e,d}$, $n^\prime_e$).

Similar results have been given and discussed previously (cf. Paper IV).
The present work provides a greatly increased sample for early-type
galaxies (S0/a and earlier) and confirms, as expected, a much lower
detection frequency for early-type galaxies as compared to spirals and
later types.  Moreover the detected emission for early-type galaxies
is seen to be predominantly compact emission.  The likely origin for
this emission is discussed in section \ref{emmech} below.

\section{Emission in cluster and field galaxies}
\label{scfcomp}

The earliest studies which compared the frequency of emission between
field and cluster galaxies were in agreement in finding, {\it for a
given galaxy morphological type}, a lower frequency of ELGs in
clusters (e.g.  Osterbrock 1960, Gisler 1978, Dressler et al. 1985,
Hill \& Oegerle 1993). However Biviano et al. (1997) identified a
hitherto unsuspected systematic effect which causes an overestimate in
the fraction of ELGs at fainter magnitudes, due to the bias that
operates against the successful determination of redshifts for faint
galaxies without emission lines. When field galaxies are on average
fainter than the cluster galaxy sample and redshift data are
incomplete, this systematic effect works to overestimate the 
frequency of emission in field as compared to cluster galaxies.

Based on the ESO Nearby Abell Cluster Survey spectral data
(5634 galaxies in the directions of 107 cluster candidates),
Biviano et al. concluded that the observed difference in frequency of
ELGs between field and cluster galaxies could be entirely accounted
for by the variation in this frequency with galaxy morphological type,
and the differing morphological mix between field and cluster
galaxies.  By inference, there is expected to be no difference in the
ELG frequency for galaxies of a given morphological type between
cluster and field environments, in disagreement with all earlier
studies.

For the present prism survey, field and cluster galaxies have been
surveyed in an identical manner to the same magnitude limit. Redshift
data are available for most ($\sim$ 96\%) of the galaxy sample, and
field and cluster samples have both been limited to galaxies with
velocities within 3$\sigma$ of the cluster mean. Field galaxies thus
comprise an approximately volume-limited sample, and both field and
cluster galaxies are expected to have similar distributions in both
apparent and absolute magnitude.  Accordingly, we expect comparative
frequencies of cluster and field ELGs determined from this survey to
be free of the bias identified by Biviano et al. as well as 
systematic effects due to any dependence of the H$\alpha$ emission
on absolute magnitude.

Using data from the prism survey, we have previously shown that for
galaxies of types Sa and later, there is indeed no difference in the
ELG frequency between field and cluster environments, in accord with
the conclusions of Biviano et al. (cf. Paper IV and references
therein). A similar result has been obtained by Gavazzi et al. (1998)
for galaxies of types Sa and later.  These authors compared H$\alpha$
E.W.s for volume-limited samples of cluster and supercluster field
galaxies to the same magnitude limits in the Coma supercluster, and
concluded that there was no significant difference between the two
samples.

\begin{table*}
\begin{minipage}{160mm}
\caption{\label{tsurvey} Cluster galaxy survey sample}

\begin{tabular}{@{}
 l@{\hspace{0.5em}}cl@{\hspace{1.0em}}r@{\hspace{2.0em}}
 l@{\hspace{0.5em}}cl@{\hspace{1.0em}}r@{\hspace{2.0em}}
 l@{\hspace{0.5em}}cl@{\hspace{2.5em}}r} \hline

\multicolumn{1}{l}{\rule[0mm]{0mm}{4mm}\hspace*{-0.65em}CGCG} & 
\multicolumn{1}{c}{\hspace*{-0.4em}UGC} & 
\multicolumn{1}{l}{\hspace*{-0.075em}Type} & 
\multicolumn{1}{r}{$v_{\sun}$\hspace*{1.5em}} &
\multicolumn{1}{l}{\hspace*{-0.65em}CGCG} & 
\multicolumn{1}{c}{\hspace*{-0.4em}UGC} & 
\multicolumn{1}{l}{\hspace*{-0.075em}Type} & 
\multicolumn{1}{r}{$v_{\sun}$\hspace*{1.5em}} &
\multicolumn{1}{l}{\hspace*{-0.65em}CGCG} & 
\multicolumn{1}{c}{\hspace*{-0.4em}UGC} & 
\multicolumn{1}{l}{\hspace*{-0.075em}Type} & 
\multicolumn{1}{r}{$v_{\sun}$} \\

&&&
\multicolumn{1}{c}{\rule[-2mm]{0mm}{4mm}
\hspace*{-2.0em}(km ${\rm s}^{-1}$)} &&&& 
\multicolumn{1}{c}{\hspace*{-2.0em}(km ${\rm s}^{-1}$)} &&&& 
\multicolumn{1}{c}{\hspace*{-2.0em}(km ${\rm s}^{-1}$)\hspace*{-1.0em}} \\ \hline

\multicolumn{4}{l}{\underline{Abell  262}}    & 522-103  & $\ldots$ & SB:...       &  4039   & 540-048  & $\ldots$ & $\ldots$     &  6387   \\
 521-075  & $\ldots$ & SA0/a:       &  4340   &\multicolumn{4}{l}{\underline{Abell  347}}    & 540-050  &   2568   & S0           &  4752   \\
 521-077  & $\ldots$ & SAB:         &  5404   & 538-041  & $\ldots$ & S0/a         &  5819   & 540-051  & $\ldots$ & S0/a         &         \\
 521-079  &   1236   & S0:          &  4733   & 538-044  &   1790   & $\ldots$     &  6002   & 540-053  & $\ldots$ & E/S0         &  4704   \\
 522-008  & $\ldots$ & S0:          &  4423   & 538-055  &   1837   & S0           &  6582   & 540-054  &   2574   & SB0/SBa      &  5015   \\
 522-009  &   1269   & E/S0         &  3855   & 538-057A &   1841   & E            &  6373   & 540-055  &   2578   & E/S0         &  5235   \\
 522-010  &   1272   & S0           &  5005   & 538-057B &   1841   & E            &  6595   & 540-056  & $\ldots$ & $\ldots$     &  5612   \\
 522-011  & $\ldots$ & $\ldots$     &  4014   & 538-060  & $\ldots$ & S0/a         &  5659   & 540-057  &   2590   & $\ldots$     &  4719   \\
 522-012  & $\ldots$ & S...         &  4041   & 538-064  &   1859   & S0/a         &  5917   & 540-059  & $\ldots$ & S0?          &  5502   \\
 522-014  &   1277   & S0/a         &  4146   & 538-065  & $\ldots$ & S0           &  5065   & 540-061  &   2598   & S0?          &  4504   \\
 522-015  &   1283   & E/S0         &  5049   & 539-016  &   1872   & E            &  4978   & 540-062  & $\ldots$ & S0:          &  5426   \\
 522-016  & $\ldots$ & $\ldots$     & 10936   & 539-017  &   1875   & E            &  5201   & 540-063  &   2606   & E            &  4888   \\
 522-017  &   1298   & S0           &  5091   & 539-018  & $\ldots$ & S0           &         & 540-066  &   2613   & S0           &  6014   \\
 503-033  & $\ldots$ & $\ldots$     &  5611   & 539-019  & $\ldots$ & S0           &  5195   & 540-068  &   2614   & S0/a         &  5252   \\
 522-022  &   1308   & E            &  5171   & 539-020  & $\ldots$ & E/S0         &  4407   & 540-072  & $\ldots$ & SA0          &  4234   \\
 522-023  & $\ldots$ & SB0:         &  5048   & 539-021  &   1878   & E            &  5766   & 540-074  & $\ldots$ & S0:          &  4969   \\
 503-040  & $\ldots$ & E:           & 10608   & 539-022  & $\ldots$ & $\ldots$     &         & 540-075  &   2624   & S0           &  5660   \\
 522-026  & $\ldots$ & SA0:         &  4877   & 539-028  & $\ldots$ & E/S0         &  5485   & 540-077  & $\ldots$ & $\ldots$     &  5864   \\
 522-027  &   1336   & S0           &  4364   & 539-031  & $\ldots$ & S0/a         &  5645   & 540-079  & $\ldots$ & S0:          &  7236   \\
 522-030  & $\ldots$ & S0:          &  4690   & 539-033  & $\ldots$ & S0/a         &         & 540-080  & $\ldots$ & S0:          &  4983   \\
 522-032  &   1339   & SB0          &  3998   & 539-034  &   1979   & S0?          &  5750   & 540-081  &   2634   & S0/a         &  5631   \\
 522-033  & $\ldots$ & S0:          &  4284   & 539-035  &   1987   & E?           &         & 540-082  & $\ldots$ & $\ldots$     &  4850   \\
 522-034B &   1343   & S0:          &  4730   & 539-037  & $\ldots$ & S0/a         &  5859   & 540-085  & $\ldots$ & E/S0         &  4370   \\
 522-034A &   1343   & S0:          &  4618   & 539-039  & $\ldots$ & S0           &         & 540-086  &   2644   & E            &  4978   \\
 522-036  & $\ldots$ & S0/a         &  5580   & 539-042  & $\ldots$ & S0           &  4885   & 540-087  & $\ldots$ & E/S0         &  6468   \\
 522-037  &   1346   & E/S0         &  5580   & 539-043  &   2006   & E            &         & 540-088  &   2651   & E            &  7536   \\
 522-039  &   1348   & E            &  4855   & 539-044  & $\ldots$ & pec          &  4920   & 540-089  & $\ldots$ & S0           &  3342   \\
 522-040  & $\ldots$ & S0:          &  3359   & 539-050  & $\ldots$ & S0           &  5018   & 540-092  &   2657   & E?           &  5059   \\
 522-043  &   1352   & S0           &  5330   & 539-054  &   2063   & S0           &  5799   & 540-095  &   2660   & E            &  4965   \\
 522-044  & $\ldots$ & S0:          &  4928   & 539-057  &   2073   & S0?          &  5165   & 540-096  & $\ldots$ & S0           &  5751   \\
 522-045  & $\ldots$ & S0:          &  4788   & 539-058  &   2074   & S0/a         &  5601   & 525-020  &   2661   & S0           &  5980   \\
 522-046  &   1353   & E/S0         &  5254   &\multicolumn{4}{l}{\underline{Abell  400}}    & 540-097  & $\ldots$ & S0/a:        &  8194   \\
 522-047  &   1358   & S0/a         &  4458   & 415-020  &   2367   & S0           &  7380   & 540-098  &   2662   & E            &  3815   \\
 522-048  & $\ldots$ & E/S0         &  4151   & 415-033  & $\ldots$ & S0:: pec     &  8116   & 540-099  & $\ldots$ & E/S0         &  5387   \\
 522-049  & $\ldots$ & Sc:          &  4679   & 415-034  & $\ldots$ & $\ldots$     &  7229   & 540-101  & $\ldots$ & E/S0         &  4500   \\
 522-052  &   1363   & S0/a         &  4968   & 415-038  & $\ldots$ & $\ldots$     &  6384   & 540-102  & $\ldots$ & S0           &  6413   \\
 522-053  & $\ldots$ & S0/a:        &  5655   & 415-040  & $\ldots$ & S0:          &  6861   & 540-104  & $\ldots$ & S0 pec       &  5066   \\
 522-054  & $\ldots$ & S...         &  4387   & 415-041B & $\ldots$ & E/S0:        &  7142   & 540-105  &   2670   & E            &  6090   \\
 522-057  & $\ldots$ & SB0/a        &  4589   & 415-041A & $\ldots$ & E/S0:        &  6641   & 540-107  &   2673   & S0           &  4266   \\
 522-061  & $\ldots$ & S0/a:        &  5033   & 415-043B & $\ldots$ & S0::         &  6097   & 540-108  & $\ldots$ & S0:          &  4300   \\
 522-064  &   1388   & E/S0 pec     &  5359   & 415-043A & $\ldots$ & S0::         &  6410   & 540-109  &   2675   & E            &  2139   \\
 522-065  & $\ldots$ & Sb:          &  5704   & 415-044  & $\ldots$ & S0:          &  7348   & 540-110  &   2676   & E/S0         &  6749   \\
 522-068  & $\ldots$ & S...         &  4785   & 415-046  & $\ldots$ & E/S0:        &  6830   & 540-111  &   2682   & E            &  4432   \\
 522-072  & $\ldots$ & S0:          &  5371   & 415-047  & $\ldots$ & E/S0:: pec   &  6333   & 540-113  & $\ldots$ & S0:          &  4411   \\
 522-076  &   1406   & S0           &  5894   & 415-049  & $\ldots$ & S0:          &  6770   & 540-116  & $\ldots$ & E/S0         &  4173   \\
 522-080  &   1415   & S0/a         &  4796   & 415-050  & $\ldots$ & S0:          &  8215   & 540-117  &   2694   & S0?          &  6585   \\
 522-083  & $\ldots$ & $\ldots$     &  4789   & 415-051  & $\ldots$ & $\ldots$     &  8581   & 540-119  &   2698   & E            &  6421   \\
 522-084  &   1434   & S0/a         &  4540   & 415-052  & $\ldots$ & SA0/a:       &  6652   & 540-120  & $\ldots$ & S0/a:        &  4788   \\
 522-085  & $\ldots$ & S0:          &  5487   &\multicolumn{4}{l}{\underline{Abell  426}}    & 540-122  &   2708   & S0           &  5394   \\
 522-087  &   1440   & E            &  4667   & 540-035  &   2528   & S0/a         &  6296   & 540-123  &   2717   & E            &  3798   \\
 522-089  & $\ldots$ & E/S0:        &  5114   & 540-038  &   2533   & E/S0         &  1592   & 541-004  &   2725   & S0           &  6192   \\
 522-091  & $\ldots$ & E/S0         &  4686   & 540-040  &   2536   & S0           &  4792   & 541-007  &   2733   & E?           &  5331   \\
 522-092  & $\ldots$ & S0/a:        &  4214   & 540-041  & $\ldots$ & S0:          &  5355   & 541-012  & $\ldots$ & $\ldots$     &  4743   \\
 522-098  & $\ldots$ & S0:          &  4817   & 540-044  &   2554   & S0?          &  2847   & 541-013  &   2752   & S0           &  4179   \\
 522-099  & $\ldots$ & SB:...       &  5334   & 540-045  & $\ldots$ & $\ldots$     &  5830   & 541-014  & $\ldots$ & S0:          &  5580   \\
 522-101  &   1475   & E            &  4209   & 540-046  &   2559   & S0           &   557   & 541-016A &   2756   & E/S0         &  5061   \\
\end{tabular}

\end{minipage}
\end{table*}

\setcounter{table}{0}
\begin{table*}
\begin{minipage}{160mm}
  \caption{continued}
\begin{tabular}{@{}
 l@{\hspace{0.5em}}cl@{\hspace{1.0em}}r@{\hspace{2.0em}}
 l@{\hspace{0.5em}}cl@{\hspace{1.0em}}r@{\hspace{2.0em}}
 l@{\hspace{0.5em}}cl@{\hspace{2.5em}}r} \hline

\multicolumn{1}{l}{\rule[0mm]{0mm}{4mm}\hspace*{-0.65em}CGCG} & 
\multicolumn{1}{c}{\hspace*{-0.4em}UGC} & 
\multicolumn{1}{l}{\hspace*{-0.075em}Type} & 
\multicolumn{1}{r}{$v_{\sun}$\hspace*{1.5em}} &
\multicolumn{1}{l}{\hspace*{-0.65em}CGCG} & 
\multicolumn{1}{c}{\hspace*{-0.4em}UGC} & 
\multicolumn{1}{l}{\hspace*{-0.075em}Type} & 
\multicolumn{1}{r}{$v_{\sun}$\hspace*{1.5em}} &
\multicolumn{1}{l}{\hspace*{-0.65em}CGCG} & 
\multicolumn{1}{c}{\hspace*{-0.4em}UGC} & 
\multicolumn{1}{l}{\hspace*{-0.075em}Type} & 
\multicolumn{1}{r}{$v_{\sun}$} \\

&&&
\multicolumn{1}{c}{\rule[-2mm]{0mm}{4mm}
\hspace*{-2.0em}(km ${\rm s}^{-1}$)} &&&& 
\multicolumn{1}{c}{\hspace*{-2.0em}(km ${\rm s}^{-1}$)} &&&& 
\multicolumn{1}{c}{\hspace*{-2.0em}(km ${\rm s}^{-1}$)\hspace*{-1.0em}} \\ \hline

\hspace*{-0.58em} \rule[0mm]{0mm}{4mm} 541-016B &   2756   & S0:          &         & 181-028  & $\ldots$ & E/S0:        & 15125   & 160-052  & $\ldots$ & $\ldots$     &  8196   \\
 541-018  &   2762   & E/S0         &  5442   & 181-029  & $\ldots$ & S0:          &  6992   & 160-053  & $\ldots$ & $\ldots$     &  7095   \\
\multicolumn{4}{l}{\underline{Abell  569}}    & 181-031  & $\ldots$ & E/S0         &  8252   & 160-056  &   8086   & S0           &  5844   \\
 234-031  & $\ldots$ & $\ldots$     &  5792   & 181-033  &   4972   & S0?          &  7075   & 160-057  & $\ldots$ & $\ldots$     &  7506   \\
 234-040  & $\ldots$ & $\ldots$     &  6047   & 181-034  & $\ldots$ & $\ldots$     &  6310   & 160-059  & $\ldots$ & $\ldots$     &  7675   \\
 234-047  &   3659   & E            &  5944   & 181-035  &   4974   & S0?          &  7023   & 160-061A & $\ldots$ & $\ldots$     &  7907   \\
 234-048  & $\ldots$ & S0/a:        &  5661   & 181-038  & $\ldots$ & $\ldots$     &  6657   & 160-061B & $\ldots$ & $\ldots$     &  6697   \\
 234-053  & $\ldots$ & S0:          &  5895   & 181-040  &   5001   & SB0          &  1687   & 160-063  & $\ldots$ & S0:          &  6023   \\
 234-054  & $\ldots$ & $\ldots$     &  6003   & 181-041  & $\ldots$ & E/S0:        & 12742   & 160-065  & $\ldots$ & E/S0         &  7145   \\
 234-058  & $\ldots$ & $\ldots$     &         &\multicolumn{4}{l}{\underline{Abell 1656}}    & 160-066  & $\ldots$ & $\ldots$     &  8004   \\
 234-059  & $\ldots$ & S0:          &  5341   & 159-098  & $\ldots$ & E/S0:        &  8010   & 160-068  &   8092   & $\ldots$     &  7660   \\
 234-064  & $\ldots$ & $\ldots$     &  5748   & 159-101  & $\ldots$ & $\ldots$     &  7745   & 160-069  & $\ldots$ & $\ldots$     &  6704   \\
 234-068  & $\ldots$ & S0/a:        &  6156   & 159-102  &   8017   & $\ldots$     &  7061   & 160-070  & $\ldots$ & E/S0         &  8430   \\
 234-070  &   3695   & E/S0         &  5795   & 159-104  & $\ldots$ & S0/a:        &  6159   & 160-071  & $\ldots$ & $\ldots$     &  5682   \\
 234-073  &   3696   & E/S0         &  6150   & 159-106  & $\ldots$ & $\ldots$     &  7945   & 160-072  & $\ldots$ & S0/a         &  5916   \\
 234-074  & $\ldots$ & S0:          &  4820   & 159-111  &   8026   & S0/a         &  7627   & 160-074  &   8097   & S0/a         &  7164   \\
 234-075  &   3699   & S0           &  5836   & 159-112  & $\ldots$ & $\ldots$     &  6413   & 160-076  & $\ldots$ & S0:          &  7972   \\
 234-077  & $\ldots$ & S0:          &  6108   & 159-113  &   8028   & E/S0         &  8365   & 160-078  & $\ldots$ & E/S0:        &  5554   \\
 234-078  & $\ldots$ & E/S0:        &  5472   & 159-114  & $\ldots$ & SB0/a:       &  7033   & 160-079  & $\ldots$ & $\ldots$     &  6336   \\
 234-080  & $\ldots$ & SB:0/a:      &  6174   & 159-115  & $\ldots$ & $\ldots$     &  6202   & 160-080  & $\ldots$ & $\ldots$     &  7088   \\
 234-081  &   3713   & E/S0         &  6500   & 159-118  &   8038   & $\ldots$     &  7863   & 160-081  & $\ldots$ & $\ldots$     &  6650   \\
 234-082  & $\ldots$ & E/S0:        &  5648   & 159-119  & $\ldots$ & $\ldots$     &  7495   & 160-082  & $\ldots$ & $\ldots$     &  7678   \\
 234-083  & $\ldots$ & E/S0:        &  5747   & 160-015  & $\ldots$ & S0:          &  7356   & 160-083  & $\ldots$ & E/S0         &  7980   \\
 234-084  & $\ldots$ & E/S0:        &  6310   & 160-016  & $\ldots$ & $\ldots$     &  7177   & 160-084  & $\ldots$ & $\ldots$     &  5675   \\
 234-085  & $\ldots$ & $\ldots$     &  6158   & 160-017  &   8049   & $\ldots$     &  6989   & 160-085  & $\ldots$ & $\ldots$     &  6812   \\
 234-086  & $\ldots$ & E/S0         &  6393   & 160-018  & $\ldots$ & $\ldots$     &  7049   & 160-086  & $\ldots$ & $\ldots$     &  4859   \\
 234-087  & $\ldots$ & S0:          &  4820   & 160-019  & $\ldots$ & $\ldots$     &  7115   & 160-087  & $\ldots$ & E/S0 pec     &  6841   \\
 234-088B &   3719   & $\ldots$     &         & 160-020  & $\ldots$ & $\ldots$     &  4968   & 160-088  & $\ldots$ & E/S0         &  7780   \\
 234-089  & $\ldots$ & $\ldots$     &  5894   & 160-021  &   8057   & E/S0         &  6915   & 160-089  &   8100   & E?           &  4670   \\
 234-091  &   3720   & E            &  5925   & 160-022  & $\ldots$ & S0:          &  6486   & 160-090  & $\ldots$ & E/S0         &  6875   \\
 234-095  & $\ldots$ & E/S0         &  6102   & 160-023  & $\ldots$ & $\ldots$     &  6883   & 160-091  & $\ldots$ & $\ldots$     &  6790   \\
 234-096  & $\ldots$ & E/S0         &  5824   & 160-024  & $\ldots$ & $\ldots$     &  7506   & 160-092  & $\ldots$ & E/S0:        &  4755   \\
 234-097  &   3725   & E/S0         &  6171   & 160-026  & $\ldots$ & S0/a:        &  7525   & 160-093  & $\ldots$ & $\ldots$     &  7209   \\
 234-098  & $\ldots$ & S0/a:        &  5132   & 160-027  & $\ldots$ & $\ldots$     &  6297   & 160-094  & $\ldots$ & E/S0:        &  6717   \\
 234-101  & $\ldots$ & E/S0:        &  5487   & 160-028  &   8065   & S0           &  7630   & 160-095  & $\ldots$ & E/S0:        &  5848   \\
 234-105  & $\ldots$ & S0:          &  6229   & 160-029  & $\ldots$ & S0/a:        &  6296   & 160-097  & $\ldots$ & $\ldots$     &  8045   \\
 234-108  & $\ldots$ & E/S0         &  5823   & 160-031  & $\ldots$ & $\ldots$     &  6849   & 160-098  &   8103   & S0           &  7224   \\
 234-110  & $\ldots$ & S0/a:        &         & 160-032  & $\ldots$ & S0/a:        &  7581   & 160-100  & $\ldots$ & $\ldots$     &  6148   \\
 234-111  &   3758   & E/S0 pec     &  5678   & 160-033  & $\ldots$ & $\ldots$     &  6273   & 160-101  & $\ldots$ & $\ldots$     &  5978   \\
 234-113  & $\ldots$ & $\ldots$     &  9988   & 160-034  & $\ldots$ & $\ldots$     &  8064   & 160-102  & $\ldots$ & $\ldots$     &  8342   \\
 234-115  & $\ldots$ & $\ldots$     &         & 160-035  & $\ldots$ & S0/a:        &  7495   & 160-103  & $\ldots$ & $\ldots$     &  8009   \\
 235-006  & $\ldots$ & S0:          &  5861   & 160-037  & $\ldots$ & $\ldots$     &  7463   & 160-104  & $\ldots$ & $\ldots$     &  6678   \\
\multicolumn{4}{l}{\underline{Abell  779}}    & 160-039  &   8070   & E            &  7362   & 160-105  & $\ldots$ & $\ldots$     &  6900   \\
 181-005  & $\ldots$ & E/S0:        & 14974   & 160-040  & $\ldots$ & $\ldots$     &  5475   & 160-106  & $\ldots$ & E/S0:        &  9401   \\
 181-008  & $\ldots$ & SB0/a        &  7171   & 160-041  & $\ldots$ & $\ldots$     &  7230   & 160-108  & $\ldots$ & $\ldots$     &  8071   \\
 181-009  & $\ldots$ & S0:          &  6634   & 160-042  & $\ldots$ & $\ldots$     &  6087   & 160-109  &   8106   & E            &  6740   \\
 181-011  & $\ldots$ & E/S0         &  7213   & 160-044A &   8072   & E            &  6775   & 160-111  & $\ldots$ & E/S0         &  6392   \\
 181-014  & $\ldots$ & $\ldots$     &  6781   & 160-044B &   8072   & E            &  6302   & 160-112  & $\ldots$ & $\ldots$     &  6568   \\
 181-015  & $\ldots$ & $\ldots$     &  7036   & 160-045  & $\ldots$ & $\ldots$     &  6356   & 160-113B & $\ldots$ & $\ldots$     &  9902   \\
 181-018  &   4939   & S0/a         &  6379   & 160-046A & $\ldots$ & $\ldots$     &  7343   & 160-114  &   8110   & E            &  6494   \\
 181-020  & $\ldots$ & $\ldots$     &  6465   & 160-046B & $\ldots$ & $\ldots$     &  7234   & 160-115  & $\ldots$ & $\ldots$     &  7268   \\
 181-021  & $\ldots$ & S0:          &  6394   & 160-047  & $\ldots$ & $\ldots$     &  6118   & 160-116  & $\ldots$ & $\ldots$     &  7268   \\
 181-022  & $\ldots$ & S0/a:        &  7062   & 160-048A & $\ldots$ & E/S0:        &  5861   & 160-117  & $\ldots$ & $\ldots$     &  6781   \\
 181-024A &   4942   & E            &  6948   & 160-048B & $\ldots$ & $\ldots$     &  6990   & 160-118  & $\ldots$ & $\ldots$     & 11114   \\
 181-024B & $\ldots$ & E/S0         &  5180   & 160-049  & $\ldots$ & $\ldots$     &  7237   & 160-119  & $\ldots$ & $\ldots$     &  5737   \\
 181-027  & $\ldots$ & $\ldots$     &  7321   & 160-051  &   8080   & S0/a         &  7410   & 160-120  & $\ldots$ & E/S0:        &  7365   \\
\end{tabular}

\end{minipage}
\end{table*}

\setcounter{table}{0}
\begin{table*}
\begin{minipage}{160mm}
  \caption{continued.}
\begin{tabular}{@{}
 l@{\hspace{0.5em}}cl@{\hspace{1.0em}}r@{\hspace{2.0em}}
 l@{\hspace{0.5em}}cl@{\hspace{1.0em}}r@{\hspace{2.0em}}
 l@{\hspace{0.5em}}cl@{\hspace{2.5em}}r} \hline

\multicolumn{1}{l}{\rule[0mm]{0mm}{4mm}\hspace*{-0.65em}CGCG} & 
\multicolumn{1}{c}{\hspace*{-0.4em}UGC} & 
\multicolumn{1}{l}{\hspace*{-0.075em}Type} & 
\multicolumn{1}{r}{$v_{\sun}$\hspace*{1.5em}} &
\multicolumn{1}{l}{\hspace*{-0.65em}CGCG} & 
\multicolumn{1}{c}{\hspace*{-0.4em}UGC} & 
\multicolumn{1}{l}{\hspace*{-0.075em}Type} & 
\multicolumn{1}{r}{$v_{\sun}$\hspace*{1.5em}} &
\multicolumn{1}{l}{\hspace*{-0.65em}CGCG} & 
\multicolumn{1}{c}{\hspace*{-0.4em}UGC} & 
\multicolumn{1}{l}{\hspace*{-0.075em}Type} & 
\multicolumn{1}{r}{$v_{\sun}$} \\

&&&
\multicolumn{1}{c}{\rule[-2mm]{0mm}{4mm}
\hspace*{-2.0em}(km ${\rm s}^{-1}$)} &&&& 
\multicolumn{1}{c}{\hspace*{-2.0em}(km ${\rm s}^{-1}$)} &&&& 
\multicolumn{1}{c}{\hspace*{-2.0em}(km ${\rm s}^{-1}$)\hspace*{-1.0em}} \\ \hline

\hspace*{-0.64em} \rule[0mm]{0mm}{4mm} 160-121B & $\ldots$ & $\ldots$     &  6371   & 160-143  & $\ldots$ & $\ldots$     &  6828   & 160-167  & $\ldots$ & $\ldots$     &  8210   \\
 160-122  & $\ldots$ & $\ldots$     &  4634   & 160-144  & $\ldots$ & $\ldots$     &  5965   & 160-168  & $\ldots$ & S0:          &  7759   \\
 160-121A & $\ldots$ & $\ldots$     &  6811   & 160-145  & $\ldots$ & $\ldots$     &  6664   & 160-169  & $\ldots$ & $\ldots$     &  5965   \\
 160-123  & $\ldots$ & $\ldots$     &  8220   & 160-146  &   8133   & S0           &  7334   & 160-170  &   8154   & S0           &  5443   \\
 160-124  & $\ldots$ & S0           &  8492   & 160-149  & $\ldots$ & E/S0:        &  5484   & 160-171  & $\ldots$ & S0/a         &  6917   \\
 160-125  & $\ldots$ & SB:0/a:      &  7208   & 160-151  &   8137   & S0           &  7387   & 160-174  & $\ldots$ & $\ldots$     &  5602   \\
 160-126  & $\ldots$ & E/S0         &  6812   & 160-152  & $\ldots$ & $\ldots$     &  7556   & 160-175  & $\ldots$ & E/S0:        &  6358   \\
 160-128  &   8117   & E/S0         &  6012   & 160-153  & $\ldots$ & $\ldots$     &  5807   & 160-176B &   8167   & $\ldots$     &         \\
 160-129  & $\ldots$ & E/S0:        &  7521   & 160-155  &   8142   & E/S0         &  7887   & 160-177  & $\ldots$ & E/S0:        &  5939   \\
 160-131  & $\ldots$ & $\ldots$     &  5441   & 160-156  & $\ldots$ & $\ldots$     &  7072   & 160-181  &   8175   & E            &  5968   \\
 160-133  & $\ldots$ & S0:          &  6363   & 160-157  & $\ldots$ & E/S0         &  7764   & 160-182  &   8178   & E            &  6909   \\
 160-134  & $\ldots$ & $\ldots$     &  7112   & 160-158  & $\ldots$ & $\ldots$     &  7188   & 160-184  & $\ldots$ & $\ldots$     &  7075   \\
 160-135  & $\ldots$ & $\ldots$     &  7997   & 160-161  & $\ldots$ & E/S0:        &  7572   & 160-187  & $\ldots$ & $\ldots$     &  7944   \\
 160-136  &   8122   & S0/a         &  6925   & 160-162  & $\ldots$ & S0/a:        &  5580   & 160-189B &   8194   & $\ldots$     &  7163   \\
 160-137  & $\ldots$ & E/S0:        &  8793   & 160-163  & $\ldots$ & $\ldots$     &  6872   & 160-190  & $\ldots$ & $\ldots$     &  7851   \\
 160-138  & $\ldots$ & E/S0:        &  6940   & 160-164B & $\ldots$ & $\ldots$     &         & 160-192  & $\ldots$ & S0/a:        &  6307   \\
 160-141  & $\ldots$ & E            &  5012   & 160-165  & $\ldots$ & E/S0         &  6211   & 160-193  & $\ldots$ & $\ldots$     &  7304   \\
 160-142  & $\ldots$ & $\ldots$     &  7665   & 160-166  & $\ldots$ & $\ldots$     &  7406   & 160-195  &   8206   & S0/a         &  6655   \\ \hline
\end{tabular}

\vspace{\baselineskip}

Notes to the Table:

A total of 24 galaxies were not surveyed due to plate defects, viz. 9
galaxies on a defocussed region of Plate 15270 for Abell 1656 (CGCG
nos. 159-106, 159-114, 159-119, 160-016, 160-029, 160-034, 160-035,
160-044A and 160-044B); 12 galaxies whose spectra were overlapped by
adjacent stellar or galaxy spectra (CGCG nos. 159-113, 522-027,
538-041, 538-057A, 538-057B, 538-060, 539-043, 415-034, 415-041A,
415-041B, 540-089 and 541-016A); and 3 galaxies which lay outside the
overlap region of the survey plate pair (CGCG nos.  503-033, 503-040
and 540-035).

Notes on double systems:

CGCG 522-034A and B: N and S components, $m_{p}$ $\sim$ 14.7 and 15.1 respectively.

CGCG 415-043A and B: NW and SE components, $m_{p}$ $\sim$ 16.3 and 16.3 respectively.

CGCG 541-016A and B: S and N components, $m_{p}$ $\sim$ 15.8 and 15.8 respectively.

CGCG 234-088A and B: S and N components, $m_{p}$ $\sim$ 15.4 and 16.6 respectively.

CGCG 181-024A and B: NE and SW components, $m_{p}$ $\sim$ 13.7 and 14.5 respectively.

CGCG 160-046A and B: N and S components, $m_{p}$ $\sim$ 15.5 and 15.9 respectively.

CGCG 160-048A and B: W and E components, $m_{p}$ $\sim$ 15.9 and 16.2 respectively.

CGCG 160-061A and B: S and N components, $m_{p}$ $\sim$ 15.8 and 16.1 respectively.

CGCG 160-113A and B: W and E components, $m_{p}$ $\sim$ 16.0 and 16.8 respectively.

CGCG 160-121A and B: S and N components, $m_{p}$ $\sim$ 15.3 and 15.7 respectively.

CGCG 160-164A and B: E and W components, $m_{p}$ $\sim$ 16.3 and 16.3 respectively.

CGCG 160-176A and B: W and E components, $m_{p}$ $\sim$ 13.5 and 15.2 respectively.

CGCG 160-189A and B: E and W components, $m_{p}$ $\sim$ 14.0 and 16.5 respectively.

Explanations of the columns in Table \ref{tsurvey}.

 \indent Column 1.  CGCG number (Zwicky et al. 1960--1968). The
 numbering of CGCG galaxies in field 160 (Abell 1656) which has a
 subfield covering the dense central region of the cluster, follows
 that of the listing of the CGCG in the SIMBAD database. The
 enumeration is in strict order of increasing Right Ascension, with
 galaxies of lower declination preceeding in cases of identical Right
 Ascension.

 \indent Column 2.  UGC number (Nilson 1973)

 \indent Column 3. Galaxy type taken from UGC or estimated from the
 PSS. 

 \indent Column 4. Heliocentric velocity taken from the NASA
 Extragalactic Database (NED).
\end{minipage}
\end{table*}

\begin{table*}
\begin{minipage}{160mm}
\caption{\label{gsurvey} Galaxies detected in  H$\alpha$ emission}

\begin{tabular*}{15cm}
{@{}l@{\extracolsep\fill \hspace{0.5em}}cr@{\hspace{0.5em}}
r@{\hspace{1.5em}}r@{\hspace{0.5em}}rcclrccc@{}} \hline

\multicolumn{1}{l}{\rule[0mm]{0mm}{4mm}\hspace*{-0.65em}CGCG} & 
\multicolumn{1}{c}{\hspace*{-0.4em}UGC} & 
\multicolumn{4}{c}{R.A. (1950) Dec. \hspace*{-2.0em}} &
\multicolumn{1}{c}{$r$} &
\multicolumn{1}{c}{$m_{p}$} &
\multicolumn{1}{l}{\hspace*{-0.075em}Type} & 
\multicolumn{1}{c}{$v_{\sun}$} &
\multicolumn{2}{c}{H$\alpha$ emission} &
\multicolumn{1}{c}{Notes} \\

&&&&&&
\multicolumn{1}{c}{\rule[-2mm]{0mm}{4mm} $(r_{A})$ } &&&
\multicolumn{1}{c}{(km ${\rm s}^{-1}$)\hspace*{-1.0em}} &
\multicolumn{1}{c}{Vis.} &
\multicolumn{1}{c}{Conc.} & \\ \hline

\multicolumn{13}{l}{\rule[0mm]{0mm}{4mm}\underline{Abell  262}}  \\
 522-011  & $\ldots$ &  1 & 46.4 &  34 & 43 &  0.8 & 15.4 & ${\rm S0}^{\dagger}$     &  4014 & MS   & C    & * \\
 522-039  &  1348    &  1 & 49.9 &  35 & 55 &  0.0 & 14.8 & E                        &  4855 & W    & N    & * \\
 522-053  & $\ldots$ &  1 & 51.0 &  36 & 23 &  0.3 & 15.4 & S0/a:                    &  5655 & W    & D    & \\
 522-072  & $\ldots$ &  1 & 53.4 &  35 & 20 &  0.5 & 15.2 & S0:                      &  5371 & M    & D    & * \\
\multicolumn{13}{l}{\underline{Abell  347}} \\
 539-031  & $\ldots$ &  2 & 24.4 &  41 & 47 &  0.2 & 15.0 & S0/a                     &  5645 & S    & VC   & * \\
 539-044  & $\ldots$ &  2 & 30.5 &  41 &  8 &  1.0 & 15.3 & pec                      &  4920 & MS   & VD   & \\
\multicolumn{13}{l}{\underline{Abell  400}} \\
 415-020  &  2367    &  2 & 51.0 &   6 &  4 &  0.9 & 15.3 & S0                       &  7380 & W:   & VD   & \\
 415-033  & $\ldots$ &  2 & 53.8 &   4 & 25 &  1.2 & 15.2 & S0:: pec                 &  8116 & M    & N    & \\
 415-038  & $\ldots$ &  2 & 54.3 &   6 &  0 &  0.2 & 15.3 & ${\rm S0}^{\dagger}$     &  6384 & S    & N    & \\
 415-043B & $\ldots$ &  2 & 55.3 &   5 & 35 &  0.2 & (16.3) & S0::                     &  6097 & W:   & D    & \\
 415-050  & $\ldots$ &  2 & 56.6 &   5 & 56 &  0.3 & 15.0 & S0:                      &  8215 & S    & N    & \\
 415-052  & $\ldots$ &  2 & 57.5 &   5 & 36 &  0.6 & 15.3 & SA0/a:                   &  6652 & W:   & N    & \\
\multicolumn{13}{l}{\underline{Abell  569}} \\
 234-085  & $\ldots$ &  7 &  7.0 &  48 & 16 &  0.3 & 15.5 & $\ldots$                 &  6158 & MW   & N    & \\
 234-096  & $\ldots$ &  7 &  7.9 &  47 &  1 &  1.2 & 15.3 & E/S0                     &  5824 & M    & C    & \\
 234-113  & $\ldots$ &  7 & 11.9 &  49 & 42 &  1.0 & 15.7 & ${\rm Sc}^{\dagger}$     &  9988 & W    & D    & \\
 234-115  & $\ldots$ &  7 & 13.0 &  49 & 58 &  1.2 & 15.7 & $\ldots$                 &       & W    & N    & \\
\multicolumn{13}{l}{\underline{Abell  1656}} \\
 159-101  & $\ldots$ & 12 & 50.3 &  27 & 40 &  1.4 & 15.3 & ${\rm Irr}^{\dagger}$    &  7745 & S    & N    & \\
 159-102  &  8017    & 12 & 50.4 &  28 & 39 &  1.3 & 14.5 & ${\rm Sab}^{\dagger}$    &  7061 & M    & D    & \\
 160-020  & $\ldots$ & 12 & 53.7 &  27 & 57 &  0.7 & 15.5 & ${\rm Sa}^{\dagger}$     &  4968 & S    & C    & \\
 160-026  & $\ldots$ & 12 & 54.1 &  27 & 33 &  0.8 & 15.5 & S0/a:                    &  7525 & MW   & N    & \\
 160-033  & $\ldots$ & 12 & 54.5 &  27 & 10 &  1.0 & 15.1 & ${\rm E}^{\dagger}$      &  6273 & MW   & N    & * \\
 160-068  &  8092    & 12 & 56.2 &  27 & 51 &  0.4 & 14.2 & ${\rm (R}^{\prime}{\rm )SA0-?}^{\dagger}$     
                                                                                     &  7660 & W    & N    & \\
 160-078  & $\ldots$ & 12 & 56.7 &  27 & 55 &  0.3 & 15.1 & E/S0:                    &  5554 & S    & N    & * \\
 160-156  & $\ldots$ & 12 & 59.6 &  28 &  3 &  0.4 & 15.4 & ${\rm SA0}^{\dagger}$    &  7072 & W    & N    & \\
 160-158  & $\ldots$ & 12 & 59.7 &  27 & 55 &  0.5 & 15.1 & ${\rm S0\hspace{0.5em}pec?}^{\dagger}$&  7188 & S    & N    & \\
 160-169  & $\ldots$ & 13 &  0.6 &  26 & 47 &  1.3 & 15.7 & ${\rm S}^{\dagger}$      &  5965 & M    & N    & \\
 160-189B &  8194    & 13 &  3.9 &  29 & 20 &  1.5 & (16.5) & S            &  7163 & W    & D    & \\
\hspace*{-0.72em} \rule[-1.2mm]{0mm}{4mm}
 160-193  & $\ldots$ & 13 &  4.8 &  28 & 18 &  1.3 & 15.5 & ${\rm Sc+}^{\dagger}$    &  7304 & S    & N    & \\ \hline
\end{tabular*}

\vspace{\baselineskip}

\noindent  

Notes on individual objects:

CGCG 522-011: = ARK 59 (Arakalian 1975). This high surface brightness
elliptical blue object is a component of a triple system and is likely
to be interacting with another component, CGCG 522-013 = V Zw 113
(``distorted blue Sc'', cf. Zwicky 1971; $v_{\sun}$ = 4019 km ${\rm
s}^{-1}$).

CGCG 522-039: = NGC 708, the brightest galaxy in Abell 262.  It has a
radio jet (Bridle \& Purley 1984), and also a double ionised gas
component, one aligned with the stars of the galaxy and the other
strongly decoupled, suggesting an external origin (Plena et al. 1998).
Nuclear emission ratio, [NII]/H$\alpha$ $\sim$ 2 (Miller \& Owen 2002).

CGCG 522-072: = V Zw 144 (``blue fuzzy elliptical disk compact'', cf.
Zwicky 1971).

CGCG 539-031: = MRK 1176. Markarian starburst galaxy paired with
companion (Keel \& van Soest 1992; Miller \& Owen 2002).

CGCG 160-033: = ARK 395 (``Compact symmetrical blue object with 
envelope'', cf. Arakalian 1975)

CGCG 160-078: = MRK 58. A blue disk galaxy which shows a very 
asymmetric gas distribution (cf. Bravo-Alfaro et al. 2000 ). Type 
given by NED is SBa.

Explanations of columns in Table \ref{gsurvey}

 \indent Columns 1 and 2. As columns 1 and 2 of Table \ref{tsurvey}.

 \indent Columns 3 and 4.  Right Ascension and Declination (1950.0) of
 the galaxy centre taken from the CGCG.

 \indent Column 5.  Radial distance in Abell radii (Abell 1958) of the
 galaxy with respect to the cluster centre. Adopted positions for the
 cluster centres and values of the Abell radii are given in Paper IV.
  
 \indent Column 6. CGCG photographic magnitude. For double galaxies,
 magnitude estimates for individual components obtained by eye from
 PSS are given in parentheses.

 \indent Column 7. As column 3 of Table \ref{tsurvey}. ${}^{\dagger}$
 indicates that types were taken from NED.

\end{minipage}
\end{table*}

\pagebreak

\begin{table*}
\begin{minipage}{160mm}

 \indent Column 8. As column 4 of Table \ref{tsurvey}.

 \indent Column 9. A visibility parameter describing how readily the
 H$\alpha$ emission is seen on the plates according to a five-point
 scale (S strong, MS medium-strong, M medium, MW medium-weak, W weak).

 \indent Column 10.  A concentration parameter describing the spatial
 distribution of the emission and contrast with the underlying
 continuum, on a five-point scale (VD very diffuse, D diffuse, N
 normal, C concentrated, VC very concentrated).

 \indent Column 11.  Notes.  An asterisk in this column indicates that
 a note on this galaxy appears below the Table.

\end{minipage}
\end{table*}

\begin{table}
 \begin{center}
  \caption{\label{etypes} Galaxies detected in emission for different type classes.}
\begin{tabular}{@{}l@{}c@{}c@{}c@{}c@{}c@{}c@{}c@{}c@{}} \hline
\multicolumn{1}{l}{\rule[0mm]{0mm}{4mm} Type} &
\multicolumn{2}{c}{Total} &
\multicolumn{2}{c}{Compact} &
\multicolumn{2}{c}{Diffuse} &
\multicolumn{2}{c}{Percentage} \\
\multicolumn{1}{l}{\rule[-2mm]{0mm}{4mm}}
 &&&
\multicolumn{2}{c}{ELGs} &
\multicolumn{2}{c}{ELGs} &
\multicolumn{2}{c}{ELGs} \\ \cline{2-3} \cline{4-5} \cline{6-7} \cline{8-9} 
\multicolumn{1}{l}{\rule[-2mm]{0mm}{4mm} \rule[0mm]{0mm}{4mm}} &
\multicolumn{1}{c}{$n_t$} &
\multicolumn{1}{c}{$n^\prime_t$} &
\multicolumn{1}{c}{$n_{e,c}$} &
\multicolumn{1}{c}{$n^\prime_{e,c}$} &
\multicolumn{1}{c}{$n_{e,d}$} &
\multicolumn{1}{c}{$n^\prime_{e,d}$} &
\multicolumn{1}{c}{$p_e$} &
\multicolumn{1}{c}{$p^\prime_e$} \\ \hline
\multicolumn{1}{l}{\rule[0mm]{0mm}{4mm}
\hspace*{-1.35em} E }          &  39 & ( 55) & 1 & ( 2) & 0 & ( 0) & 2.6 & ( 3.6) \\
 E-S0         &  82 & ( 84) & 2 & ( 2) & 0 & ( 0) & 2.4 & ( 2.4) \\
 S0           & 127 & (173) & 3 & ( 8) & 2 & ( 2) & 3.9 & ( 5.8) \\
 S0/a         &  60 & ( 67) & 5 & ( 5) & 1 & ( 1) &10.0 & ( 9.0) \\
 Sa           &  55 & ( 58) & 8 & ( 9) & 8 & ( 8) &29.1 & (29.3) \\
 Sab          &  25 & ( 27) & 3 & ( 3) & 7 & ( 8) &40.0 & (40.7) \\
 Sb           &  40 & ( 40) &10 & (10) & 8 & ( 8) &45.0 & (45.0) \\
 Sbc,Sc       &  44 & ( 45) & 7 & ( 8) &11 & (11) &40.9 & (42.2) \\
 Sc--Irr,Irr  &  11 & ( 13) & 1 & ( 2) & 3 & ( 3) &36.4 & (38.5) \\
 pec          &  21 & ( 21) &14 & (14) & 2 & ( 2) &76.2 & (76.2) \\
 S$\ldots$    &  61 & ( 69) &12 & (13) &13 & (13) &41.0 & (37.7) \\
\multicolumn{1}{l}{\rule[-2mm]{0mm}{4mm} \hspace*{-1.35em}
 $\ldots$  }  & 105 & ( 18) &12 & ( 2) & 1 & ( 0) &12.4 & (11.1) \\ \hline
\end{tabular}
\end{center}
Note to the Table: $n^\prime_t$ $n^\prime_{e,c}$ $n^\prime_{e,d}$ 
and $p^\prime_e$ for the various samples are obtained by including NED types, 
where available, for galaxies with indeterminate type.
\end{table}

Using the prism survey data in this paper, we now compare ELG
frequencies for E,S0,S0/a galaxies between field and cluster
environments.  We adopt the definitions of projected radial distance
from the cluster centre, $R$; local surface density, $\Sigma$; and
cluster type, $CT$ given in Paper IV. For the latter parameter, $CT$,
cluster galaxies were taken as those surveyed galaxies with $r \le
1.0r_{\rm A}$; field galaxies were either those surveyed galaxies with
$r > 1.5r_{\rm A}$, or (for Abell 262, 347, 400 and 779) with $r >
1.0r_{\rm A}$. Clusters were ranked according to increasing mean space
density of galaxies in the central regions of the cluster ($r \le
0.5r_{\rm A}$) with the first rank for field galaxies.  (For further
discussion, cf. Paper IV).

If the frequency of ELGs varies systematically from field to cluster,
we expect this frequency to show a dependence on one or more of the
three parameters, $R$, $\Sigma$ and $CT$.  A Kendall rank test shows
no significant correlation between the fraction of E,S0,S0/a galaxies
detected in emission and each of $R$, $\Sigma$ and $CT$ (significance
levels of 0.0$\sigma$, -1.1$\sigma$, and -0.8$\sigma$
respectively). Thus we confirm that there is no dependence of the
frequency of ELGs among early type galaxies on either local surface
density or cluster environment in accord with results of Biviano et
al.

A Kendall rank test also shows no significant correlation between the
fraction of E,S0,S0/a galaxies detected in {\it compact} emission and
each of $R$, $\Sigma$ and $CT$ (significance levels of -0.3$\sigma$,
-0.3$\sigma$, and 0.1$\sigma$ respectively). This result is expected
from the previous one, because most ELGs in the E,S0,S0/a sample have
compact emission.  However this result is in contrast to
the previous finding (cf. Paper IV) of an enhancement of compact
emission for cluster spirals.

In previous work, we have suggested that the enhancement of compact
emission in cluster galaxies of types Sa and later is due to
circumnuclear starbursts triggered by tidal interactions associated
with sub-cluster merging and on-going processes of virialisation.
With this scenario, one expects an enhancement of compact emission in
the non-virialised later-type galaxy population.  The degree of
enhancement is likely to be related to the strengths of the varying
gravitational fields in clusters of differing galaxy density and stage
of relaxation.

If this scenario is correct, any corresponding enhancement of compact
emission in early-type cluster galaxies may be less evident.  As a
cluster continues to form, and spirals are transformed into earlier
type galaxies, any enhancement of circumnuclear starburst emission in
early-type galaxies is likely to be masked by the increasing number
of these galaxies in the cluster.

Accordingly, we attempt to test whether there is an enhancement of
starburst emission in early-type galaxies in the clusters
surveyed, using an alternative method described in the next section.

\section{Enhancement of starburst emission in cluster early-type galaxies}
\label{emmech}

In previous work, restricted to a sample of the surveyed cluster
galaxies of types Sa and later, we have shown that compact emission
can convincingly be identified as due to circumnuclear starburst
emission, whereas diffuse emission originates from more normal star
formation in the disks of the spiral galaxies.  The compact emission
was shown to be associated both with a tidal disturbance of the galaxy
and also with the presence of a bar. For spirals in this sample, a
strong association was also found between compact emission and the presence of
a tidal companion, which suggests that much of this emission is caused by
circumnuclear starbursts associated with local galaxy--galaxy
interactions.  However galaxies classified as peculiar show no
tendency to have tidal companions, although a very high percentage
($\sim$ 76 per cent) of these galaxies show compact emission (see
Table \ref{etypes}).  As discussed in Paper IV, a natural
explanation of this latter result is that peculiars are predominantly
on-going mergers, in which the companion is already indistinguishable
from its merger partner, and the compact emission arises from the
starburst induced by the merger.

However for early-type galaxies, the association between compact
emission and star formation appears less likely. For example, in the
Palomar spectroscopic survey of nearby galactic nuclei (PSSN; Ho et
al. 1997), not a single elliptical galaxy shows emission attributable
to star formation.

The above suggests that we might detect any enhancement of starburst
emission in the early-type cluster galaxies by comparison of the
fraction of detected early-type cluster ELGs which show HII emission,
with the same fraction for a comparable field sample.  We proceed to
make this comparison as follows.  First, we determine the {\it
expected} fraction of early-type ELGs with HII emission from data for
the field from the PSSN.  This is done in section \ref{eem}.  Then, in
section \ref{oem}, we make a comparison between this expected fraction
and the {\it observed} fraction for cluster early-type ELGs, using a
compendium of published data. As will be seen, this comparison gives a
modest indication of an enhancement of starburst emission in cluster
early-type galaxies.  Furthermore we show how this enhancement of emission
can readily be explained by the effect of gravitational tidal interactions.

\subsection{Expected emission from field data}
\label{eem}

The PSSN is based on high quality optical spectra of moderate resolution for
the nucleii (r $\la$ 200 pc) of almost every bright galaxy 
(${\rm B}_{T}$ $\le$ 12.5) in the northern sky ($\delta$ $>$ $0^{o}$).
Using standard nebular diagnostics, the probable ionisation mechanisms of 
the emission-line objects in the survey have been identified.  These
spectral classifications can be used to infer the likely origin of the
compact emission detected in early-type galaxies by the present
objective prism survey in the following way.

\begin{figure}
\centering
\includegraphics[width=0.41\textwidth,angle=-90,bb=66 71 540 676]{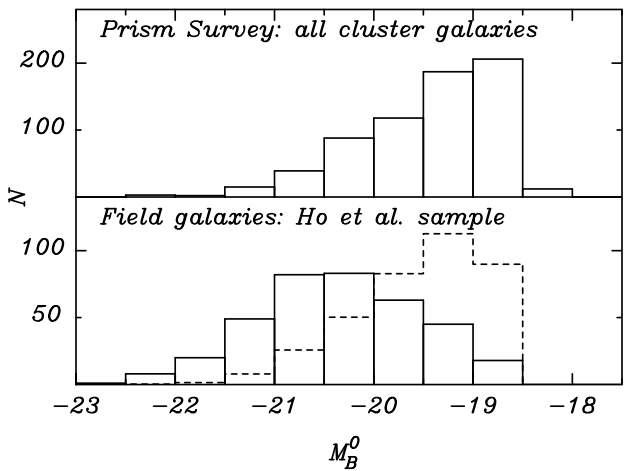}

\caption{\label{mdistribs} Distribution in absolute magnitude, 
$M_{B}^{0}$, for galaxies in the prism survey (top) and in a sub-sample
of the Palomar spectroscopic survey of nearby galactic nuclei (PSSN)
restricted to $M_{B}^{0}$ $\le$ -18.5 (bottom). 
The dashed histogram shows the volume weighted distribution for the PSSN
sub-sample.}
\end{figure}

First, we select a sub-sample of the PSSN sample whose distribution in
absolute magnitude approximately matches that of the CGCG galaxies in
the prism survey. This match is important because the ionisation
mechanism for emission-line objects in the PSSN has a strong
dependence on absolute magnitude.  In Figure \ref{mdistribs}, we show
the distributions of absolute asymptotic B magnitudes corrected for
internal and Galactic extinction, $M_{B}^{0}$, for galaxies in the
prism survey and for a sub-sample of the PSSN restricted to
$M_{B}^{0}$ $\le$ -18.5.  Values of $M_{B}^{0}$ for the prism survey
galaxies were derived from CGCG magnitudes, $m_{p}$, converted to the
$B_{T}$ system following Paturel, Bottinelli
\& Gouguenheim (1994) and corrected for galactic and internal
absorption following Sandage \& Tammann (1987).  Values of $M_{B}^{0}$
for the PSSN sub-sample were taken from Ho et
al. (1997).\footnote{Although details of the derivations of
$M_{B}^{0}$ for the two samples differ, the resulting magnitude
differences are not considered significant in the present context.}
In both cases absolute magnitudes were derived assuming ${\rm H}_{0}$
= 75 km ${\rm s}^{-1}$ ${\rm Mpc}^{-1}$.

Whereas the PSSN is an apparent magnitude-limited sample, the prism
survey sample is approximately volume limited and thus has a relative
preponderance of fainter galaxies.  To match the samples more closely,
the PSSN sub-sample was weighted according to the volume surveyed.
The volume weighted distribution is shown by the dashed histogram in
Figure \ref{mdistribs}.

\begin{figure}
\centering
\includegraphics[width=0.5\textwidth]{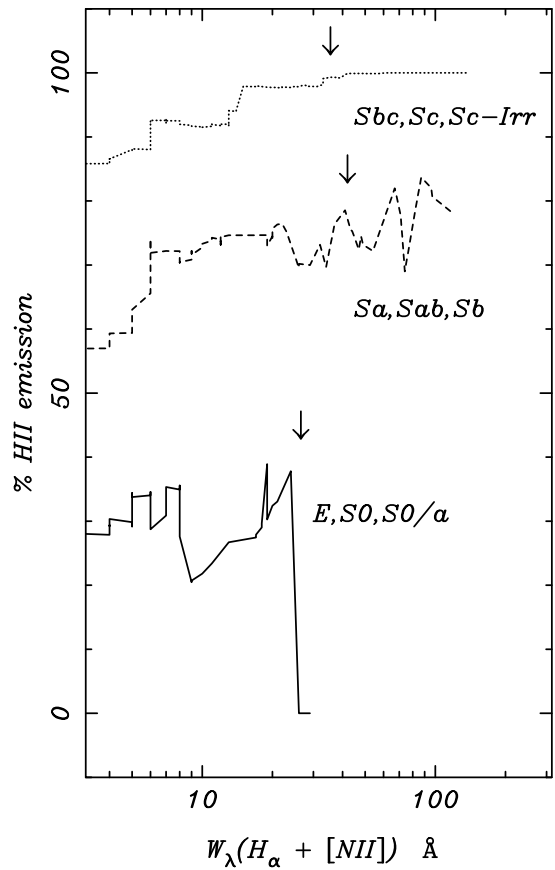}

\caption{\label{ph2} Cumulative percentages of the volume weighted PSSN
field galaxy sample (Ho et al. 1997) with HII emission, plotted
against the combined equivalent width, $W_{\lambda}({\rm H}\alpha +
{\rm [NII]})$.  The cumulative percentages (with the integrated number
of galaxies, $n$, increasing from right to left) are for all galaxies
with equivalent width $\ge$ $W_{\lambda}$. Plots are given for each of
the type groups, E,S0,S0/a ($n \ge 3$); Sa,Sab,Sb ($n \ge 3$); and
Sbc,Sc,Sc--Irr ($n \ge 10$).  Arrows indicate the values of $W_{\lambda}$
at which the fraction of galaxies with equivalent width $\ge$
$W_{\lambda}$ equals the fraction of galaxies detected in compact
emission by the cluster prism survey.}
\end{figure}

Next, we have used the PSSN volume weighted sample to determine
the percentages of galaxies which have HII emission for each 
of the three type groups, E,S0,S0/a; Sa,Sab,Sb; and Sbc,Sc,Sc--Irr.
These percentages are shown in Figure \ref{ph2}.  In this Figure, the 
percentage of galaxies with HII emission is plotted against the
combined equivalent width of H$\alpha$ and [NII] ($\lambda$ 6584\AA),
$W_{\lambda}({\rm H}\alpha + {\rm [NII]})$.  The percentage plotted
is a cumulative percentage for all galaxies with equivalent width
$\ge$ $W_{\lambda}$.

Values of $W_{\lambda}$ plotted in Figure \ref{ph2} from the PSSN are for
nuclear emission from slit spectroscopy. By contrast, ELG detections
for the prism survey are based on integrated global emission.
However, for strong emission ($W_{\lambda}$ $\ge$ 10\AA), we expect
equivalent widths determined by slit spectroscopy and integrated
global emission to be well correlated (cf. Kennicutt \& Kent
1983). Accordingly, for $W_{\lambda}$ $\ge$ 10\AA, we can assume that
percentages of galaxies with HII emission in Figure \ref{ph2} are
representative of emission galaxies detected in the prism survey.

Arrows in Figure \ref{ph2} indicate the values of $W_{\lambda}$ at which the
fraction of galaxies with equivalent width $\ge$ $W_{\lambda}$ equals
the fraction of galaxies detected in compact emission by the prism
survey. Percentage values for each type group at these points thus
indicate the percentages of prism survey compact ELGs which are
expected to have HII emission.

It is seen that a majority of late type compact ELGs are expected to
have HII emission ($\sim$ 75\% for types Sa,Sab,Sb; and $\sim$ 100\%
for types Sbc,Sc,Sc--Irr).  This is in agreement with previous work where
it is assumed that compact emission from galaxy types Sa and later is
predominantly due to star formation (cf. Papers II-IV).  By contrast,
only $\sim$ 30\% of compact ELGs of early type (E,S0,S0/a) are expected
to have HII emission.  The remaining early type compact ELGs
are expected to have AGN or LINER emission.

\begin{table}
\begin{center}
\caption{\label{spt} Spectral classifications for galaxies with 
compact emission}

\begin{tabular}{@{}l@{\hspace{3.0em}}l@{\hspace{-1.0em}}cc} \hline

\multicolumn{1}{l}{\rule[0mm]{0mm}{4mm}\hspace*{-0.65em}CGCG} &
\multicolumn{1}{l}{\hspace*{-0.8em}Type} &
\multicolumn{1}{c}{Spectral type} &
\multicolumn{1}{l}{\rule[-2mm]{0mm}{4mm} Ref.} \\ \hline

\multicolumn{3}{l}{\rule[0mm]{0mm}{4mm}\underline{Abell 262}}  & \\
 521-074 & S: pec       & HII            & 2 \\
 522-003 & pec          & HII            & 2 \\
 522-011 & ${\rm S0}^{\dagger}$          & HII            & 2 \\
 522-020 & SBb          & Sbrst          & 2 \\
 522-039 & E            & Sy2            & 2 \\
 522-058 & SBa          & Sbrst          & 2 \\
 522-077 & SBb: pec     & HII            & 2 \\
\multicolumn{3}{l}{\underline{Abell 347}}  & \\
 538-043 & pec          & SF             & 1 \\
 539-024 & SBb          & HII            & 3 \\
 539-025 & SB pec       & SF             & 1 \\
 539-031 & S0/a         & SF             & 1 \\
\multicolumn{3}{l}{\underline{Abell 400}}  & \\
 415-050 & S0:          & Mix            & 1 \\
\multicolumn{3}{l}{\underline{Abell 426}}  & \\
 540-064 & SBb          & Sy2            & 2 \\
 540-094 & Sbc          & HII:           & 2 \\
 540-103 & pec:         & Sy2            & 2 \\
 541-009 & SBc          & AGN            & 2 \\
\multicolumn{3}{l}{\underline{Abell 569}}  & \\
 234-056 & S pec        & HII            & 2 \\
 234-057 & pec          & SF             & 1 \\
 234-066 & pec:         & SF             & 1 \\
 234-071 & SB: pec      & AGN            & 2 \\
 234-079A& S: pec       & AGN            & 2 \\
 234-115 & ...          & SF             & 1 \\
\multicolumn{3}{l}{\underline{Abell 779}}  & \\
 181-023 & S            & Ab+            & 1 \\
 181-030 & SB:b         & AGN:           & 2 \\
\multicolumn{3}{l}{\underline{Abell 1367}} & \\
 \hspace*{0.15em}  97-026 & SBa pec      & SF             & 1 \\
 \hspace*{0.15em}  97-068 & SBc pec      & SF             & 1 \\
 \hspace*{0.15em}  97-079 & S:...pec     & SF             & 1 \\
 \hspace*{0.15em}  97-087 & Sd pec       & SF             & 1 \\
 \hspace*{0.15em}  97-114 & S0/a: pec    & HII            & 2 \\
 \hspace*{0.15em}  97-125 & Sa: pec:     & SF             & 1 \\
 126-110 & Sab pec      & Sy1            & 2 \\
 127-049 & SBab         & SF             & 1 \\
 127-052 & SA0          & Lin            & 1 \\
 127-055 & SAa          & SF             & 1 \\
 127-071 & S... pec     & HII            & 2 \\
 127-095 & SBb          & HII            & 2 \\
\multicolumn{3}{l}{\underline{Abell 1656}} & \\
 159-101 & ${\rm Irr}^{\dagger}$         & HII            & 2 \\
 160-020 & ${\rm Sa}^{\dagger}$          & HII            & 2 \\
 160-026 & S0/a:        & HII            & 2 \\
 160-055 & SB:ab        & SF             & 1 \\
 160-064 & pec          & Sbrst          & 2 \\
 160-067 & pec          & SF             & 1 \\
 160-068 & (${\rm R}^{\prime}$)${\rm SA0-?}^{\dagger}$      & 
 SF             & 1 \\
 160-075 & pec          & Sbrst          & 2 \\
 160-078 & E/S0:        & Sbrst          & 2 \\
 160-127 & pec      & SF             & 1 \\
 160-130 & pec:         & SF             & 1 \\
 160-148A& S pec       & Mix            & 1 \\
 160-156 & ${\rm SA0}^{\dagger}$          & Sy             & 2 \\
 160-158 & ${\rm S0\: pec?}^{\dagger}$      & Sbrst          & 2 \\
\end{tabular} 
\end{center}
\end{table}

\setcounter{table}{3}
\begin{table}
\begin{center}
\caption{continued.}

\begin{tabular}{@{}l@{\hspace{3.0em}}l@{\hspace{-1.0em}}cc} \hline

\multicolumn{1}{l}{\rule[0mm]{0mm}{4mm}\hspace*{-0.65em}CGCG} &
\multicolumn{1}{l}{\hspace*{-0.8em}Type} &
\multicolumn{1}{c}{Spectral type} &
\multicolumn{1}{l}{\rule[-2mm]{0mm}{4mm} Ref.} \\ \hline

 160-160 & pec          & HII            & 2 \\
 160-179 & S: pec       & HII            & 2 \\
 160-180 & pec          & HII            & 2 \\
\rule[-2mm]{0mm}{4mm}
 \hspace*{-0.625em} 160-193  & S:...        & Sbrst        & 2 \\ \hline
\end{tabular} 
\end{center}
Explanation of the columns of the Table.

Column 1. CGCG number (Zwicky et al. 1960--1968).  See notes to column 1 of
Table \ref{tsurvey} above.

Column 2. Galaxy type taken from papers III \& IV, and this paper. ${}^{\dagger}$
indicates that types were taken from NED.

Columns 3 \& 4: Spectral classification taken from: 1. Miller \& Owen
(2002): SF -- star-forming galaxy; Ab+ -- predominantly
absorption-line spectrum, although with slight emission of [NII] and
sometimes [SII]; Sey -- Seyfert; Lin -- LINER; Mix -- nuclear spectrum
that of a Seyfert or LINER, with off-nuclear spectrum showing star
formation. 2. NED: either NED galaxy classification (Sbrst, Sy, Sy1, Sy2)
or classification as HII or AGN emission from the line ratios of
emission lines in the UZC Spectral Archive (Faldo et al. 1999).
\end{table}

\subsection{Comparison of expected and observed HII emission for early-type
cluster galaxies}
\label{oem}

The above expected percentages of compact ELGs of different galaxy
types with HII emission may be compared with actual percentages from
spectroscopic data. In Table \ref{spt}, we list spectral
classifications for 54 compact ELGs in the 8 clusters surveyed.  These
represent 69\% of the total of 78 compact ELGs which have a
velocity within 3$\sigma$ of the cluster mean.  Spectral
classifications were taken from Miller \& Owen (2002) or derived from
data given in NED. For these latter, the galaxy classification was
either taken from NED, or, when not available, classification of the
emission as either HII or AGN was made using emission line
ratios in the UZC Spectral Archive (Faldo et al. 1999), following the
method of Veilleux and Osterbrock (1987).

\begin{table}
\begin{center}
  \caption{\label{percents} Percentages of galaxies with compact
  HII emission in clusters and the field.}
\begin{tabular}{lrrrcrrr} \hline

\multicolumn{1}{l}{\rule[0mm]{0mm}{4mm} Type} &
\multicolumn{3}{c}{Cluster} &&
\multicolumn{3}{c}{Field} \\ \cline{2-4} \cline{6-8}  

\multicolumn{1}{l}{\rule[0mm]{0mm}{4mm}} &
\multicolumn{1}{c}{Total} &
\multicolumn{2}{c}{\% compact} &&
\multicolumn{1}{c}{Total} &
\multicolumn{2}{c}{\% compact} \\ 

 && 
\multicolumn{2}{c}{HII emission} &&&
\multicolumn{2}{c}{HII emission} \\ 

\multicolumn{1}{l}{\rule[-2mm]{0mm}{4mm}} & 
\multicolumn{1}{c}{$n_{c}$} &  
\multicolumn{1}{c}{$p_{ce,c}$} &  
\multicolumn{1}{c}{$p_{ce,c}^{d}$} &&  
\multicolumn{1}{c}{$n_{f}$} &  
\multicolumn{1}{c}{$p_{ce,f}$} &  
\multicolumn{1}{c}{$p_{ce,f}^{d}$} \\ \hline

\multicolumn{1}{l}{\rule[0mm]{0mm}{4mm} \hspace*{-1.35em} E,S0,S0/a} 
 & 302 & 3.6 & 1.3 && 28 & 7.1 & 0.0 \\
\multicolumn{1}{l}{\rule[-2mm]{0mm}{4mm} \hspace*{-1.35em} Sa + later} 
 & 177 & 19.2 & 7.3 && 45 & 8.9 & 0.0 \\ \hline

\end{tabular}
\end{center}
Note to the Table: $p_{ce,c}$ $p_{ce,c}^{d}$ and $p_{ce,f}$ $p_{ce,f}^{d}$ 
are the percentages of galaxies with compact HII emission and which both have
compact HII emission and are disturbed, for cluster and field galaxies
respectively.  NED types for galaxies with otherwise indeterminate type
have been included where available.
\end{table}

For the three type groups, E,S0,S0/a; Sa,Sab,Sb; and Sbc,Sc,Sc--Irr,
restricted to galaxies within one Abell radius of the cluster centres
(or within 0.5 Abell radii of each of the two subcluster centres in
Abell 569), the observed percentages of galaxies with HII emission are
68\% ($n=11$), 80\% ($n=10$), and 100\% ($n=3$) respectively.  For the
two later type groups, these values are in good agreement with the
expected percentages of galaxies with HII emission.  For galaxies
classified as peculiar, the observed percentage is 90\% ($n=10$) in
accord with the suggestion above that these galaxies are predominantly
on-going mergers.  However for early type galaxies, the observed
percentage is higher than expected (significance level $\sim$
2.8$\sigma$), and suggests an enhancement of HII emission for the
early-type cluster galaxies with compact emission.

If this enhancement of HII emission in early-type cluster galaxies is
real, what is its likely cause? Obviously, the simplest explanation is
that previously suggested for cluster later types, viz. the effects of
gravitational tidal interactions on the galaxies. For the present
sample of cluster early-type galaxies (types E, S0 and S0/a),
restricted to galaxies within one Abell radius of the cluster centres,
there is a very strong correlation between the incidence of compact
HII emission and a disturbed morphology for the galaxy (significance
level of 8.2$\sigma$).  Four out of nine disturbed galaxies show
compact HII emission, while only 6 out of 270 undisturbed galaxies
show this emission.  This is a similar result to that found previously
for cluster galaxies of types Sa and later, and indeed suggests that
any enhancement of HII emission in cluster early-type spirals is due
to gravitational tidal effects as is the case for later type cluster
galaxies.

In Table \ref{percents} we compare the fractions of galaxies in the
field and in clusters which have compact HII emission ($p_{ce,f}$ and
$p_{ce,c}$ respectively) and the fractions which have compact HII
emission and in addition are tidally disturbed ($p^{d}_{ce,f}$ and
$p^{d}_{ce,c}$ respectively).  Cluster galaxies are defined to be
those within $1.0{\rm r}_{A}$ of the cluster centres (or within
$0.5{\rm r}_{A}$ of each of the two subcluster centres in Abell 569),
while field galaxies are defined to be those outside $1.0{\rm r}_{A}$
for the clusters Abell 262, 347, 400, 569 and 779, and outside
$1.5{\rm r}_{A}$ for Abell 426, 1367 and 1656 (cf. Paper IV).  We make
the comparison for E,S0,S0/a and Sa + later galaxies separately.

First consider the later types.  We find that the increase in the 
percentage of galaxies with compact HII emission in clusters is
largely accounted for by an additional population of {\it disturbed}
galaxies with compact HII emission -- indeed there are {\it no} such 
galaxies in the field.  This restates our earlier findings in Paper IV 
which focussed specifically on the later types.

For the early-type galaxies, again there is a population of {\it
disturbed} galaxies with compact HII emission which have no
counterparts in the field.  The number of these galaxies ($n=4$) is
exactly that required to explain the observed excess of HII emission
in cluster early type ELGs as determined from a comparison with the
PSSN spectroscopic data for field galaxies (cf. section \ref{emmech}
above). Thus for early-types as well as later types, an enhancement of
compact HII emission in cluster galaxies is readily explained by
gravitational tidal interactions associated with cluster
virialisation.\footnote{In Table \ref{percents}, the percentage of early-type
galaxies with compact HII emission in the field is numerically greater
than in the cluster.  However, as discussed above in section \ref{scfcomp},
this cluster--field difference is not statistically significant.}
 
\section{Conclusions}
\label{concl}

In a series of papers (cf. Moss et al. 1988; Moss \& Whittle 1993;
Moss et al. 1998; Moss \& Whittle 2000) we have undertaken an
H$\alpha$ survey of an essentially complete sample of 748 CGCG
galaxies in 8 low-redshift clusters (viz. Abell 262, 347, 400, 426,
569, 779, 1367 and 1656).  This paper has presented previously
unpublished data for 383 mainly early-type galaxies and completes
publication of data for the survey.  The combined survey data show
that emission detection increases as expected from earlier to later
galaxy types (3\% for E,E--S0; 6\% for S0; 9\% for S0/a; and 41\% for
types Sa and later).

A comparison of cluster and field early-type galaxies shows a similar
frequency of emission detection.  Together with the same result
obtained for cluster and field galaxies of types Sa and later, these
data confirm the inference of Biviano et al. (1997) that differences
between the frequency of ELGs between clusters and the field can
entirely be accounted for by the differing mix of galaxy morphological
types in the two environments, while {\it for a given morphological
type} there is no difference in the frequency of ELGs between clusters
and the field.  As noted by Biviano et al., this result is in
disagreement with all or most previous studies.  It is to be noted
that this result has been obtained for galaxies with relatively strong
emission ($W_{\lambda} \ge 20$\AA). Work is in progress to extend
these results to fainter limits in equivalent width (cf. Sakai,
Kennicutt \& Moss 2001).

Although the incidence of emission does not vary between cluster and
field environments, the survey has shown that the type of emission
does vary. Detected emission is classified as `compact' or `diffuse',
identified as circumnuclear starburst or AGN emission and disk
emission respectively.  In previous work, we have shown that for
galaxies of types Sa and later, there is an enhancement of compact HII
emission in cluster galaxies as compared to the field.  This type of
emission has been shown to be strongly correlated with a tidally
disturbed morphology of the galaxy.

In the present work, a comparison of spectroscopic data for the
cluster early-type ELGs with that for field galaxies from the PSSN (Ho
et al. 1997) gives a modest indication (significance level $\sim$
2.8$\sigma$) for enhanced compact HII emission in
early-type cluster ELGs as compared to the field.  Moreover, the
compact HII emission in the early-type cluster galaxies is strongly
correlated with a disturbed galaxy morphology (significance level of
8.2$\sigma$).

For both early-type and later types, there is a population of
disturbed galaxies with compact HII emission which have no
counterparts in the field. This suggests that for cluster galaxies of
all types, enhancement of compact HII emission can readily be
explained as an enhancement of circumnuclear starburst emission due to
gravitational tidal effects.  As discussed previously (cf. Moss \&
Whittle 2000) these gravitational tidal effects are most likely to be
associated with sub-cluster merging and other processes of on-going
cluster virialisation.

\vspace{\baselineskip}

\noindent {\bf\large Acknowledgements}

\vspace{\baselineskip}

Observations were made with the Burrell Schmidt telescope of the Warner and
Swasey Observatory, Case Western Reserve University.  This research has made
use of the NASA/IPAC Extragalactic Database (NED), which is operated by the
Jet Propulsion Laboratory, California Institute of Technology, under contract
with the National Aeronautics and Space Administration.

\end{document}